\documentclass[aps,prc,reprint,superscriptaddress,showpacs]{revtex4-1}

\usepackage{graphicx}

\begin{document}

\title{\boldmath Muon spin rotation and relaxation in Pr$_{1-x}$Nd$_x$Os$_4$Sb$_{12}$: Paramagnetic states}

\author{P.-C. Ho}
\affiliation{Department of Physics, California State University, Fresno, California 93740, USA}
\author{D.~E. MacLaughlin}
\affiliation{Department of Physics and Astronomy, University of California, Riverside, California 92521, USA}
\author{Lei Shu}
\affiliation{State Key Laboratory of Surface Physics, Department of Physics, Fudan University, Shanghai 200433, People's Republic of China}
\author{O.~O. Bernal}
\affiliation{Department of Physics and Astronomy, California State University, Los Angeles, California 90032, USA}
\author{Songrui Zhao}
\altaffiliation{Current address: Department of Electrical and Computer Engineering, McGill University, Montreal, Quebec, Canada H3A~0E9.}
\affiliation{Department of Physics and Astronomy, University of California, Riverside, California 92521, USA}
\author{A.~A. Dooraghi}
\author{T. Yanagisawa}
\altaffiliation{Current address: Faculty of Science, Hokkaido University, Sapporo, Hokkaido 060-0810, Japan.}
\author{M.~B. Maple}
\affiliation{Department of Physics, University of California, San Diego, La Jolla, California 92093, USA}
\author{R.~H. Fukuda}
\affiliation{Department of Physics, California State University, Fresno, California 93740, USA}

\date{\today}

\begin{abstract}
Positive-muon ($\mu^+$) Knight shifts have been measured in the paramagnetic states of Pr$_{1-x}$Nd$_x$Os$_4$Sb$_{12}$ alloys, where $x = 0$, 0.25, 0.45, 0.50, 0.55, 0.75, and 1.00. In Pr-substituted NdOs$_4$Sb$_{12}$ ($x \le 0.75$), but not in NdOs$_4$Sb$_{12}$, Clogston-Jaccarino plots of $\mu^+$ Knight shift~$K$ versus magnetic susceptibility~$\chi$ exhibit an anomalous saturation of $K(\chi)$ at $\sim-0.5$\% for large susceptibilities (low temperatures), indicating a reduction of the coupling strength between $\mu^+$ spins and $4f$ paramagnetism for temperatures $\lesssim 15$~K\@. We speculate that itinerant Pr$^{3+}$ quadrupolar excitations, invoked to mediate the superconducting Cooper-pair interaction, might modify the $\mu^+$-$4f$ ion indirect spin-spin interaction.
\end{abstract}

\maketitle

\section{\label{sec:intro} INTRODUCTION}

The filled-skutterudite intermetallic compound PrOs$_4$Sb$_{12}$ possesses a number of unusual properties, including heavy-fermion behavior in the absence of a magnetic $4f$ ground state, a high-field ordered phase at low temperatures, unconventional superconductivity, and itinerant Pr$^{3+}$-ion quadrupolar fluctuations that may mediate Cooper pairing~\cite{[{For reviews, see }] MFHY06,*ATSK07,*MBHH09}. Isostructural NdOs$_4$Sb$_{12}$ is a ferromagnet with Curie temperature~$T_C = 0.8$~K~\cite{SSNS03,HYBF05,MHYH07}. The lattice parameters of PrOs$_4$Sb$_{12}$ and NdOs$_4$Sb$_{12}$ are nearly identical, and there is good solid solubility across the alloy series~\cite{HYYD11}. Thus Pr$_{1-x}$Nd$_x$Os$_4$Sb$_{12}$ is well suited for studying the interplay between magnetism and the unconventional properties of PrOs$_4$Sb$_{12}$. 

In a previous paper~\cite{MHSB14}, referred to hereinafter as I, we reported results of a muon spin rotation and relaxation ($\mu$SR) study of the ground states of Pr$_{1-x}$Nd$_x$Os$_4$Sb$_{12}$ alloys. The present paper presents measurements of positive-muon ($\mu^+$) Knight shifts in the paramagnetic states of Pr$_{1-x}$Nd$_x$Os$_4$Sb$_{12}$, where $x = 0$, 0.25, 0.45, 0.50, 0.55, 0.75, and 1.00. $\mu$SR is a magnetic resonance probe of magnetic behavior on the atomic scale, and our results complement an earlier study of the bulk thermal and magnetic properties of this alloy system~\cite{HYYD11}. In both I and this paper we report evidence that the coupling strength between the $\mu^+$ spin and surrounding $4f$ magnetism is anomalously suppressed in Pr-containing materials at low temperatures but not in NdOs$_4$Sb$_{12}$.

The Knight shift of a probe spin (muon or nucleus) in a paramagnet~\footnote{Although ``Knight shift'' strictly denotes only the frequency shift due to conduction-electron paramagnetism in metals, we follow the wide general usage of this term for the shift in any paramagnetic material.} is the fractional displacement of the probe-spin precession frequency in an applied magnetic field from its value \textit{in vacuo}~\cite{CBK77,Sche85,YaDdR11}. The probe spin interacts with its magnetic neighborhood via two mechanisms, dipolar coupling and the contact hyperfine interaction; the latter requires electron spin polarization density at the $\mu^+$ site. Both mechanisms contribute to the Knight shift~$K$: $K = K_\mathrm{dip} + K_c$, where $K_\mathrm{dip}$ and $K_c$ are the dipolar and contact contributions, respectively. In a paramagnetic metal with local moments, the contact interaction and the local-moment--conduction-electron exchange interaction give rise to indirect RKKY coupling. 

Dipolar shifts are nonzero only for noncubic probe-spin sites. They depend on the direction of the applied field, and in general are different for structurally equivalent probe-spin sites that are inequivalent in the field. In a single crystal this gives rise to multiple precession frequencies (generally two for a probe-spin site of axial symmetry, three for lower symmetry) that vary with orientation of the crystal. In a cubic crystal such as PrOs$_4$Sb$_{12}$ their average vanishes, leaving only the hyperfine contact contribution to the average shift. The dipolar shifts contribute to the width of the frequency spectrum, however. Oscillations associated with all the shifted frequencies are damped if the shifts are distributed, as in a disordered material or a powder with random crystallite orientations.

The Knight shift is due to the paramagnetism of the host material, and is therefore closely related to its bulk susceptibility~$\chi~$\cite{[{For a review of NMR Knight shifts in metals, see }] CBK77}. For a simple system with only one class of magnetically active electrons, $\chi$ and the Knight shift~$K$ are linearly related: $K = A\chi$, where $A$ is a coupling constant. Then if $\chi$ depends on temperature, a plot of $K(T)$ versus $\chi(T)$, with temperature~$T$ an implicit parameter (the so-called Clogston-Jaccarino plot~\cite{ClJa61}), is a straight line with zero intercept. If other contributions to $K$ and $\chi$ are temperature-independent, the $K$-$\chi$ relation remains linear but generally with a nonzero intercept. 

The $K$-$\chi$ relation can become nonlinear in a number of ways~\cite{MacL85}: 
\begin{itemize}
\item If there are two or more paramagnetic subsystems with distinguishable contributions~$\chi_i$ to $\chi$ [$\chi(T) = \sum_i \chi_i(T)$], coupled to the muon spin via separate coupling constants~$A_i$, then $K(T) = \sum_i K_i(T) = \sum_i A_i\chi_i(T)$. In this case $\chi(T)$ and $K(T)$ can have very different temperature dependencies, and $K(\chi)$ is no longer linear~\cite{CBK77}. 
\item The coupling constant is itself temperature-dependent, in which case $K(\chi)$ contains information on the coupling mechanism. Nonlinear $K(\chi)$ observed in lanthanide compounds has been attributed to temperature-dependent populations of crystalline-electric-field (CEF) split $4f$ states with different couplings to neighboring probe spins~\cite{MPL81,MacL85}. Moment instability in intermediate-valent $4f$ compounds modifies the coupling~\cite{MacL85}, as does a crossover between incoherent and coherent behavior in heavy-fermion systems~\cite{KiCo98,CYSP04}. 
\item In the case of the $\mu^+$ Knight shift, host magnetism in the neighborhood of the muon, specifically CEF splitting of $4f$ states, can be perturbed by the $\mu^+$ electric field. This modifies the susceptibility of $4f$ near neighbors, particularly if they are non-Kramers ions~\cite{FAGS94,TAGG97}, so that $K$ no longer varies linearly with the bulk susceptibility.
\end{itemize}
These and related phenomena have been discussed in a previous report of $\mu^+$ Knight shift measurements in PrOs$_4$Sb$_{12}$~\cite{SMBH09}. The $\mu$SR technique is summarized in I, and further details can be found in a number of monographs and review articles~\cite{Sche85,Brew03,ScGy95,Blun99,LKC99,YaDdR11}.

\section{\label{sec:exp}EXPERIMENT}

\begin{figure}[ht] % Fig. 1
\includegraphics[clip=,width=8.6cm]{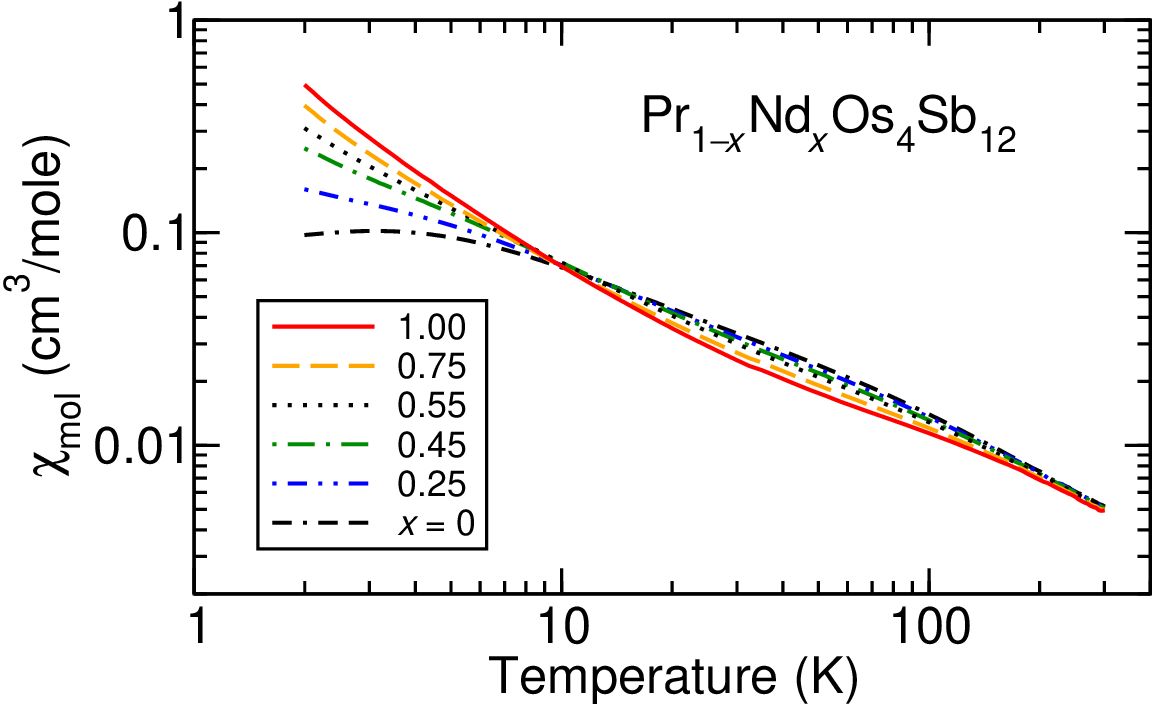}
\caption{\label{fig:PNOSchivsT}(Color online) Temperature dependence of molar magnetic susceptibility~$\chi_\mathrm{mol}$ in Pr$_{1-x}$Nd$_x$Os$_4$Sb$_{12}$, applied field 500~Oe.} 
\end{figure}

The samples for these experiments, the same as those used for zero- and low-field $\mu$SR measurements reported in I, are polycrystalline mosaics mounted on silver plates with GE varnish. Sample preparation and characterization are described in I and elsewhere~\cite{HYYD11}. The magnetic susceptibilities of all samples were measured using a Quantum Design MPMS magnetometer in a field of 500 Oe over the temperature range~2--300~K\@. Figure~\ref{fig:PNOSchivsT} gives the temperature dependence of the molar susceptibility~$\chi_\mathrm{mol}$ for several samples.
For $T \gtrsim 200$~K $\chi_\mathrm{mol}$ is nearly independent of $x$, reflecting the fact that the Hund's-rule effective magneton numbers~$p_\mathrm{eff}$ for Pr$^{3+}$ and Nd$^{3+}$ ions are almost the same (3.58 and 3.62, respectively). For temperatures below the maximum CEF splittings ($\sim$200~K and 350~K for PrOs$_4$Sb$_{12}$~\cite{MFHY06} and NdOs$_4$Sb$_{12}$~\cite{KTII08}, respectively), the details of $\chi_\mathrm{mol}(T)$ reflect CEF-split state populations. The upturns at low temperatures for Nd-rich samples are critical divergences associated with ferromagnetic transitions below 1~K.

Transverse-field $\mu$SR (TF-$\mu$SR) experiments were carried out at the M15 and M20 beam lines of the TRIUMF accelerator facility, Vancouver, Canada, in the temperature range 2--300~K, with magnetic fields~$\mathbf{H}_0$ between 3.5 and 10~kOe~\footnote{Susceptibilities were confirmed to be independent of field up to at least 10~kOe at 2~K, allowing comparison of shifts and susceptibilities measured at different fields.} applied perpendicular to the initial $\mu^+$ spin. The data were analyzed using the Paul Scherrer Institute \textsc{musrfit} fitting package~\cite{SuWo12}. 

A time-differential $\mu$SR experiment yields a ``time spectrum''~$A(t)$, which is the dependence of the asymmetry~$A$ in positron count rate on time~$t$ after muon implantation. Time spectra often contain contributions from muons that miss the sample and stop nearby in the apparatus, in addition to muons that stop in the sample. The sample environment (cold finger, mounting plate, etc.) is usually made of silver, for good thermal contact and because internal fields in pure Ag are weak and the $\mu$SR signal from it is simple~\cite{Sche85}. Then the TF-$\mu^+$SR time spectrum is generally of the form
\begin{eqnarray} \label{eq:asy} % Eq. (1)
A(t) & = & A_0[(1-f_\mathrm{Ag})G_s(t)\cos(\omega_s t + \phi) \nonumber \\
& & \qquad + f_\mathrm{Ag}\cos(\omega_\mathrm{Ag} t + \phi)],
\end{eqnarray} 
where $A_0$ is the initial asymmetry (spectrometer-dependent but normally $\sim$0.2), $f_\mathrm{Ag}$ is the fractional amplitude of the Ag component, $\phi$ is a phase factor, and $\omega_s$ and $\omega_\mathrm{Ag}$ are $\mu^+$ precession frequencies in the sample and Ag, respectively. The function~$G_s(t)$ describes damping of the precession in the sample, which is usually dominated by inhomogeneity in the precession frequency. 

Knight shift measurements in PrOs$_4$Sb$_{12}$ were carried out in a standard $\mu$SR spectrometer that yields time spectra of the form of Eq.~(\ref{eq:asy}); the Ag precession frequency is used as a reference. The $\mu^+$ Knight shift in silver is small (94~ppm)~\cite{Sche85} compared to shifts in Pr$_{1-x}$Nd$_x$Os$_4$Sb$_{12}$. We fit the data using Eq.~(\ref{eq:asy}) and two forms for $G_s(t)$: the Gaussian $\exp(-\frac{1}{2}\sigma^2t^2)$, and the phenomenological ``power exponential''~$\exp[-(\lambda t)^\beta]$, where the power~$\beta$ controls the shape of the function. For all data reported here the fit value of $\beta$ is close to 2, so that Gaussian and power-exponential fits are statistically equivalent. 

\begin{figure}[ht] % Fig. 2
\includegraphics[clip=,width=8.6cm]{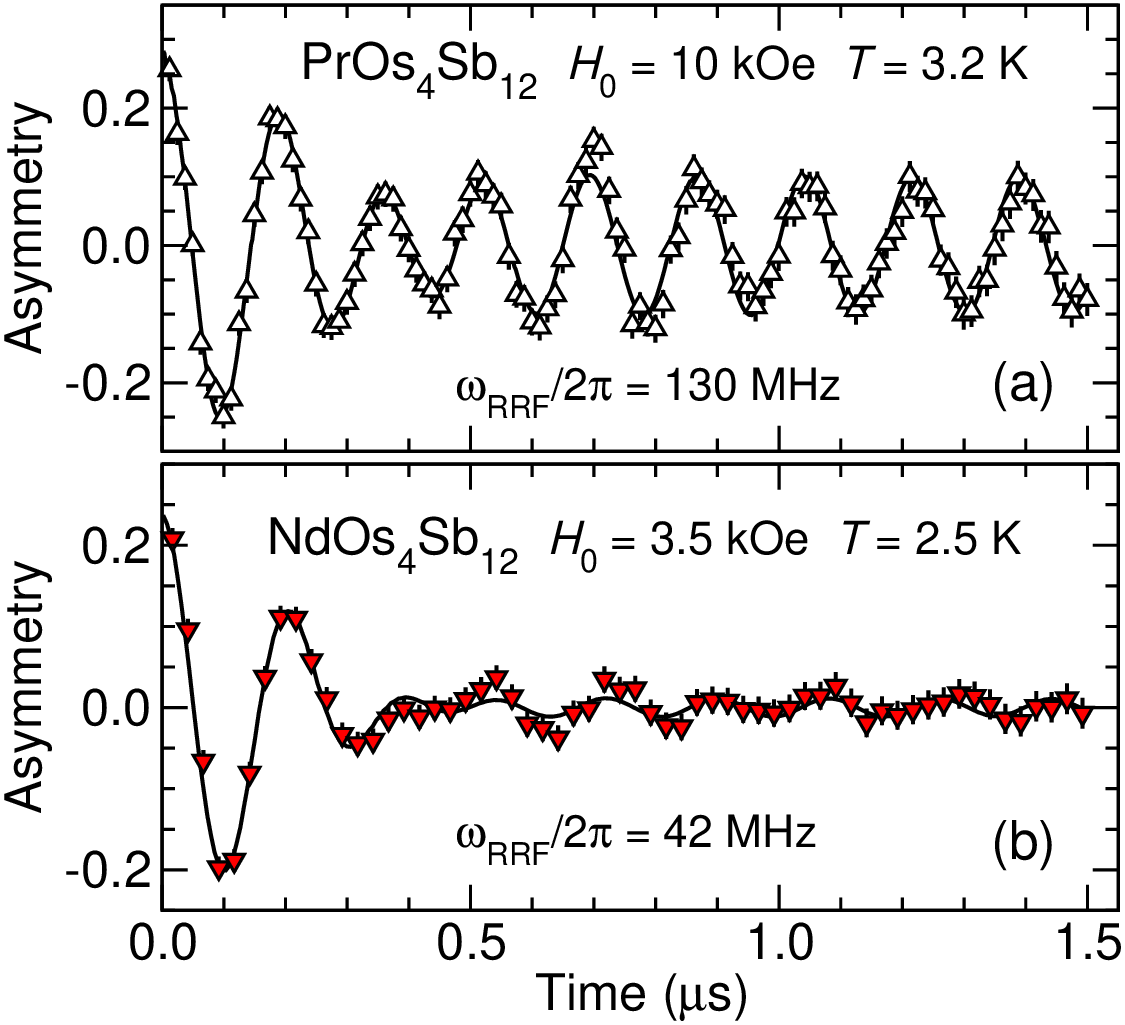}
\caption{\label{fig:POSasy}(Color online) $\mu^+$ time spectra displayed in rotating reference frames (rotation frequencies~$\omega_\mathrm{RRF}$)\@. (a)~PrOs$_4$Sb$_{12}$, single-signal $\mu$SR spectrometer insert. (b)~NdOs$_4$Sb$_{12}$, $\mu\mathrm{SR}{\times}2$ spectrometer insert, sample channel. Curves: fits of Eq.~(\protect\ref{eq:asy}) to the data.} 
\end{figure}

A representative time spectrum from PrOs$_4$Sb$_{12}$ is shown in Fig.~\ref{fig:POSasy}(a).
It is convenient to display the spectrum in a rotating reference frame (RRF)~\cite{RiBr91}, so that the observed frequencies are relatively slow beats between the precession frequencies and the RRF frequency~$\omega_\mathrm{RRF}$. The rapidly relaxing sample signal and nonrelaxing Ag component are easily distinguished, and are of comparable amplitudes.

Much of the Ag component is often due to muons that stop in the silver plate very near the sample, or in the interstices between crystallites, and hence experience a fringe field due to the sample magnetization. This reduces the Ag precession frequency relative to the reference value~$\omega_\mathrm{ref}$ due to the applied field only. A corrected value of $\omega_\mathrm{ref}$ was obtained by plotting $\omega_\mathrm{Ag}(T)$ versus $\chi(T)$, with temperature an implicit parameter, and extrapolating linearly to $\chi = 0$. The raw fractional frequency shift is then given by $K_\mathrm{raw} = \omega_s/\omega_\mathrm{ref} - 1$. 

For all samples except PrOs$_4$Sb$_{12}$, a specialized ``$\mu\mathrm{SR}{\times}2$'' spectrometer insert~\cite{CDKM97} was used. In it the reference signal is obtained from muons stopping in a flat silver ring centered on the sample. Signals from the sample and the reference have separate logical signatures, and are collected in separate histograms. A time spectrum from NdOs$_4$Sb$_{12}$ obtained using this insert is shown in Fig.~\ref{fig:POSasy}(b). The nonrelaxing signal is much weaker ($\sim$5\% of the total) but nonvanishing, so that the two-signal form of Eq.~(\ref{eq:asy}) was also used to fit data from this sample and the Nd-doped alloys. 

Substantial corrections are required for contributions to $K_\mathrm{raw}$ by the Lorentz and demagnetization fields of the sample~\cite{IHA61,SchGr65,CBK77}. These can be characterized by an effective macroscopic shift term~$K_\mathrm{Ld}$, given by
\begin{eqnarray} % Eq. (2)
K_\mathrm{Ld} & = & 4\pi\left({\textstyle\frac{1}{3}} - D_\mathrm{eff}\right)\chi_V \nonumber \\
& = & A_\mathrm{Ld}\,\chi_\mathrm{mol},\quad A_\mathrm{Ld} = 4\pi\left({\textstyle\frac{1}{3}} - D_\mathrm{eff}\right)/V_\mathrm{mol},
\end{eqnarray}
where $\chi_V$ and $\chi_\mathrm{mol}$ are the volume and molar magnetic susceptibilities, respectively, and $V_\mathrm{mol}$ is the molar volume. The effective demagnetization factor~$D_\mathrm{eff}$ is determined by the overall shape of the sample (in this case a rectangular slab~\cite{AkGa92}), the field orientation and, in the case of a powder or mosaic sample, the demagnetization factors of crystallites or grains and their volume filling fraction. To within errors $A_\mathrm{Ld} = -0.024(5)~\mathrm{mole/cm}^3$ for all samples. Inhomogeneity in the demagnetization field due to the nonellipsoidal geometry~\cite{AkGa92} is calculated to be about 11\% of the total macroscopic shift. The corrected Knight shift~$K_\mathrm{corr}$ is determined by subtracting $K_\mathrm{Ld}$ from $K_\mathrm{raw}$.

\begin{figure}[ht] % Fig. 3
\includegraphics[clip=,width=8.6cm]{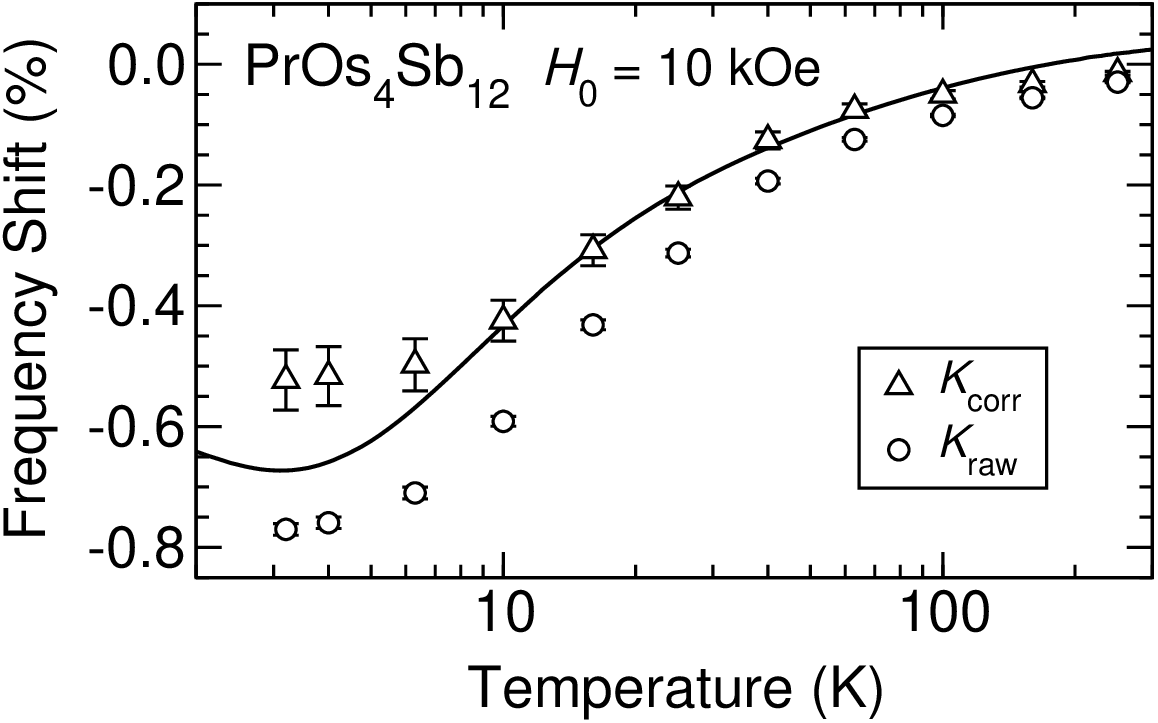}
\caption{\label{fig:POSKvsT} Temperature dependence of fractional $\mu^+$ frequency shifts in PrOs$_4$Sb$_{12}$. $K_\mathrm{raw}$: raw shift. $K_\mathrm{corr}$: Knight shift, corrected for Lorentz and demagnetization fields. Curve: $K_\mathrm{lin} (\%) = 0.06 - 7.2\chi_\mathrm{mol}(T)\ (\mathrm{cm}^3$/mole) (cf.\ Sec.~\protect\ref{sec:POSNOS}).} 
\end{figure}

An example of raw and corrected shifts in PrOs$_4$Sb$_{12}$ is shown in Fig.~\ref{fig:POSKvsT}.
The correction is a substantial fraction of the raw shift, and the error in $K_\mathrm{corr}$ is dominated by uncertainty in the geometrical factors that enter $D_\mathrm{eff}$. This error is systematic, since the same value of $A_\mathrm{Ld}$ is used for each point, so that the error bars in Fig.~\ref{fig:POSKvsT} and subsequent figures mark limits on the temperature dependence of $K_\mathrm{corr}$ rather than random uncertainties.

$\mu^+$ shift measurements in PrOs$_4$Sb$_{12}$~\cite{HSKO07,SMBH09} and Pr$_{0.6}$La$_{0.4}$Os$_4$Sb$_{12}$~\cite{HAOI07} have been reported previously, the latter for $T \le 1.5$~K\@. The results in this paper are from reanalyses of earlier data~\protect\cite{SMBH09}, and the raw shifts are in good agreement with those of Ref.~\cite{HSKO07}. In that work, however, the correction for macroscopic shifts was not reported, and an observed signal with zero raw shift (i.e., with no macroscopic shift) is unlikely to have originated from the sample.

\section{\label{sec:results} RESULTS}

\subsection{\label{sec:POSNOS} {\boldmath NdOs$_4$Sb$_{12}$} and \boldmath PrOs$_4$Sb$_{12}$}

\begin{figure}[ht] % Fig. 4
\includegraphics[clip=,width=8.6cm]{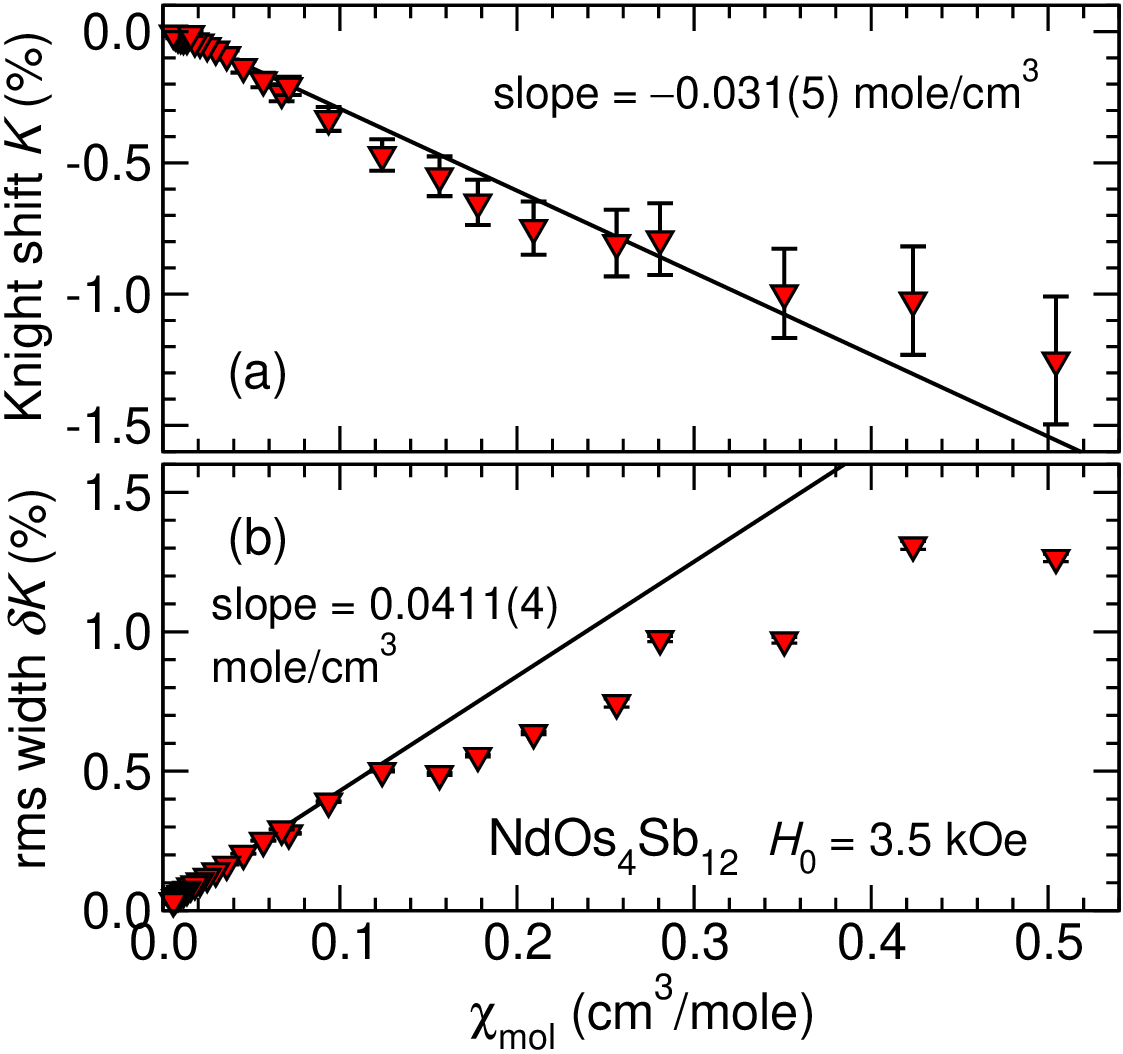}
\caption{\label{fig:NOSK+deltaKvschi}(Color online) (a)~Clogston-Jaccarino plot of $\mu^+$ Knight shift~$K$ vs molar susceptibility~$\chi_\mathrm{mol}$ in NdOs$_4$Sb$_{12}$. Line: linear fit for $\chi_\mathrm{mol}$ in the range~0.01--$0.05~\mathrm{cm}^3$/mole. (b)~rms width~$\delta K$ of Knight shift distribution vs $\chi_\mathrm{mol}$. Lines: linear fits to points for $\chi_\mathrm{mol}$ between 0.01 and $0.05~\mathrm{cm}^3$/mole.} 
\end{figure}

Figure~\ref{fig:NOSK+deltaKvschi}(a) gives a Clogston-Jaccarino plot of $\mu^+$ Knight shifts in NdOs$_4$Sb$_{12}$.
(Hereinafter $K_\mathrm{corr}$ will be referred to simply as $K$.) The overall shift is negative, i.e., the contact spin polarization at the $\mu^+$ site is antiparallel to the applied field. A good fit of the linear relation~$K = K_0 + A_c\chi_\mathrm{mol}$ to the data of Fig.~\ref{fig:NOSK+deltaKvschi}(a) is found for $\chi_\mathrm{mol}$ between 0.01 and $0.05~\mathrm{cm}^3$/mole ($130~K \gtrsim T \gtrsim 15$~K), yielding $K_0 = 0.019(7)\%$ and $A_c = -0.031(5)~\mathrm{mole/cm}^3$. For $\chi_\mathrm{mol} \gtrsim 0.05~\mathrm{cm}^3$/mole $K(\chi_\mathrm{mol})$ deviates from linearity.

The rms width~$\delta K$ of the $\mu^+$ Knight shift distribution, shown in Fig.~\ref{fig:NOSK+deltaKvschi}(b), is obtained from the relaxation rate~$\sigma$ [Eq.~(\ref{eq:asy})]: $\delta K = \sigma/\omega_\mathrm{ref}$, assuming that the spread of precession frequencies is due to a spread in Knight shifts. This assumption is indicated by the strong dependence of $\delta K$ on $\chi_\mathrm{mol}$. As for $K(\chi_\mathrm{mol})$, $\delta K(\chi_\mathrm{mol}) \propto \chi_\mathrm{mol}$ for small $\chi_\mathrm{mol}$. It can be seen that $\delta K$ and $|K|$ are comparable, indicating strong disorder. The calculated rms spread $0.0192\ \mathrm{mole/cm}^3$ in dipolar coupling constants for PrOs$_4$Sb$_{12}$~\cite{SMBH09} is less than half the initial slope $[d(\delta K)/d\chi_\mathrm{mol}]_0 = 0.0411(4)$~mole/cm$^3$ in NdOs$_4$Sb$_{12}$ [Fig.~\ref{fig:NOSK+deltaKvschi}(b)] (the lattice constants, and hence the calculated dipolar coupling constants, are nearly identical in PrOs$_4$Sb$_{12}$ and NdOs$_4$Sb$_{12}$). Although dipolar fields contribute to the width (cf.\ Sec.~\ref{sec:intro}), the considerable disorder in either the coupling constants or $\chi_\mathrm{mol}$ distributes local fields at $\mu^+$ sites and prevents observation of multiple discrete frequencies in the time spectrum (Fig.~\ref{fig:POSasy}). Disorder in the distribution of spontaneous $\mu^+$ local fields in NdOs$_4$Sb$_{12}$ below the ferromagnetic Curie temperature is reported in I, also with a width of the order of the average field.

\begin{figure}[ht] % Fig. 5
\includegraphics[clip=,width=8.6cm]{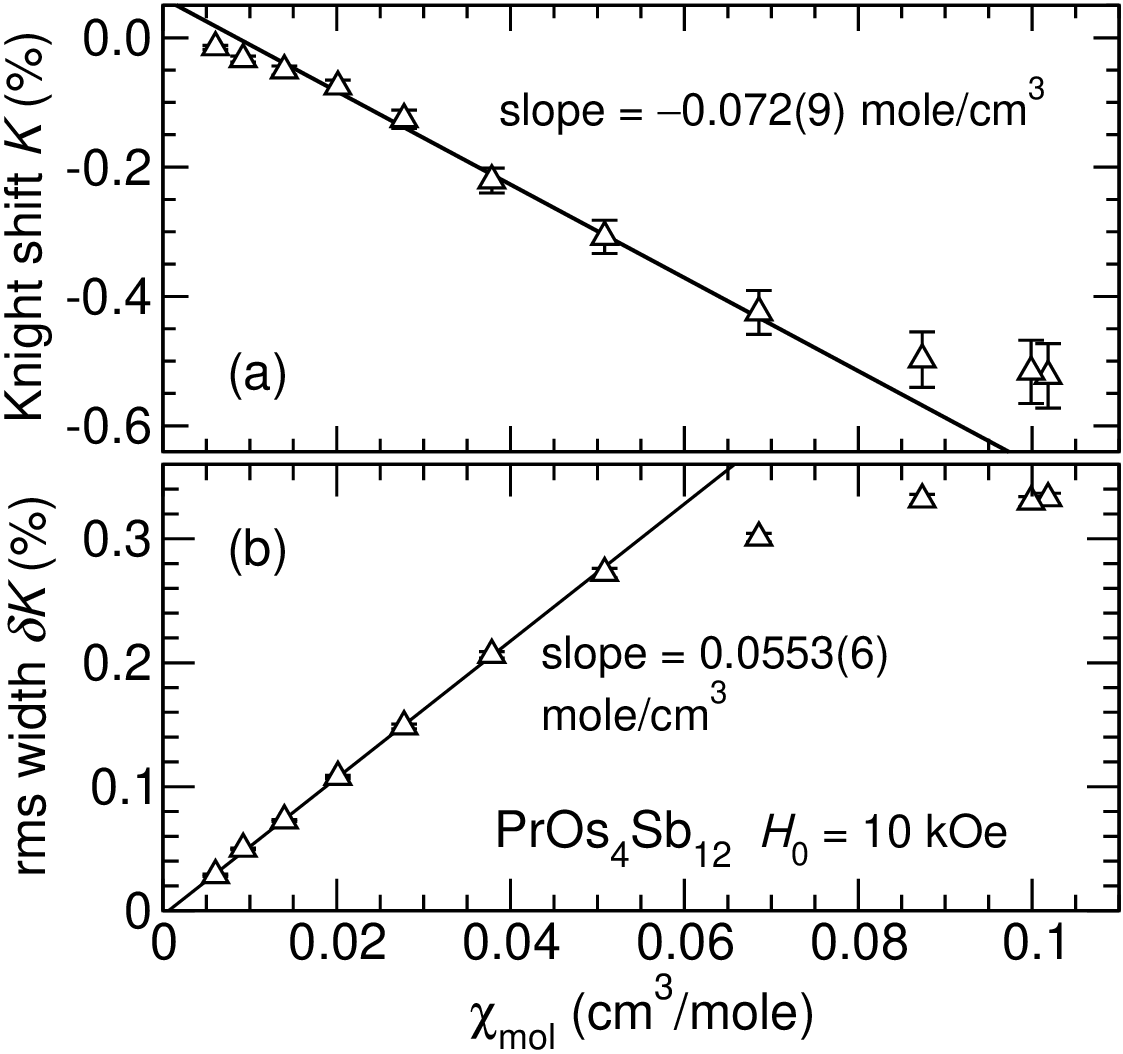}
\caption{\label{fig:POSK+deltaKvschi} (a)~Clogston-Jaccarino plot of $\mu^+$ Knight shift~$K$ vs molar susceptibility~$\chi_\mathrm{mol}$ in PrOs$_4$Sb$_{12}$. Line: linear fit for $\chi_\mathrm{mol}$ in the range~0.01--$0.05~\mathrm{cm}^3$/mole. (b)~rms width~$\delta K$ of Knight shift distribution vs $\chi_\mathrm{mol}$. Lines: linear fits to points for $\chi_\mathrm{mol}$ between 0.01 and $0.05~\mathrm{cm}^3$/mole.} 
\end{figure}

Figure~\ref{fig:POSK+deltaKvschi}(a) is a Clogston-Jaccarino plot of $\mu^+$ Knight shifts in PrOs$_4$Sb$_{12}$, $H_0 = 10$~kOe. 
For $\chi_\mathrm{mol}$ between 0.01 and $0.05~\mathrm{cm}^3$/mole, $K(\chi_\mathrm{mol})$ is again found to be linear. Here, however, the slope of the linear fit is more than a factor of two larger in magnitude than in NdOs$_4$Sb$_{12}$. The curve in Fig.~\ref{fig:POSKvsT} shows the temperature dependence of this linear fit, i.e., $K(T) = K_0 - A_c\chi_\mathrm{mol}(T)$, with $K_0 = 0.06(2)\%$ and $A_c = -0.072(9)~\mathrm{mole/cm}^3$.

In rare-earth compounds the $4f$ ions normally dominate the susceptibility, and nonlinearity in $K(\chi_\mathrm{mol})$ from multiple species of magnetic electrons is unlikely. In PrOs$_4$Sb$_{12}$ the effect of $\mu^+$ charge on the Pr$^{3+}$ CEF splitting is found to be small~\cite{SMBH09}, and Nd$^{3+}$ in NdOs$_4$Sb$_{12}$ is a Kramers ion with a ground-state moment, unlikely to be sensitive to small changes in CEF\@. The slight nonlinearity at small $\chi_\mathrm{mol}$ in both compounds is probably due to population of higher-lying CEF states with increasing temperature.

\subsection{\label{sec:alloys} \boldmath Pr$_{1-x}$Nd$_x$Os$_4$Sb$_{12}$}

\begin{figure}[ht] % Fig. 6
\includegraphics[clip=,width=8.6cm]{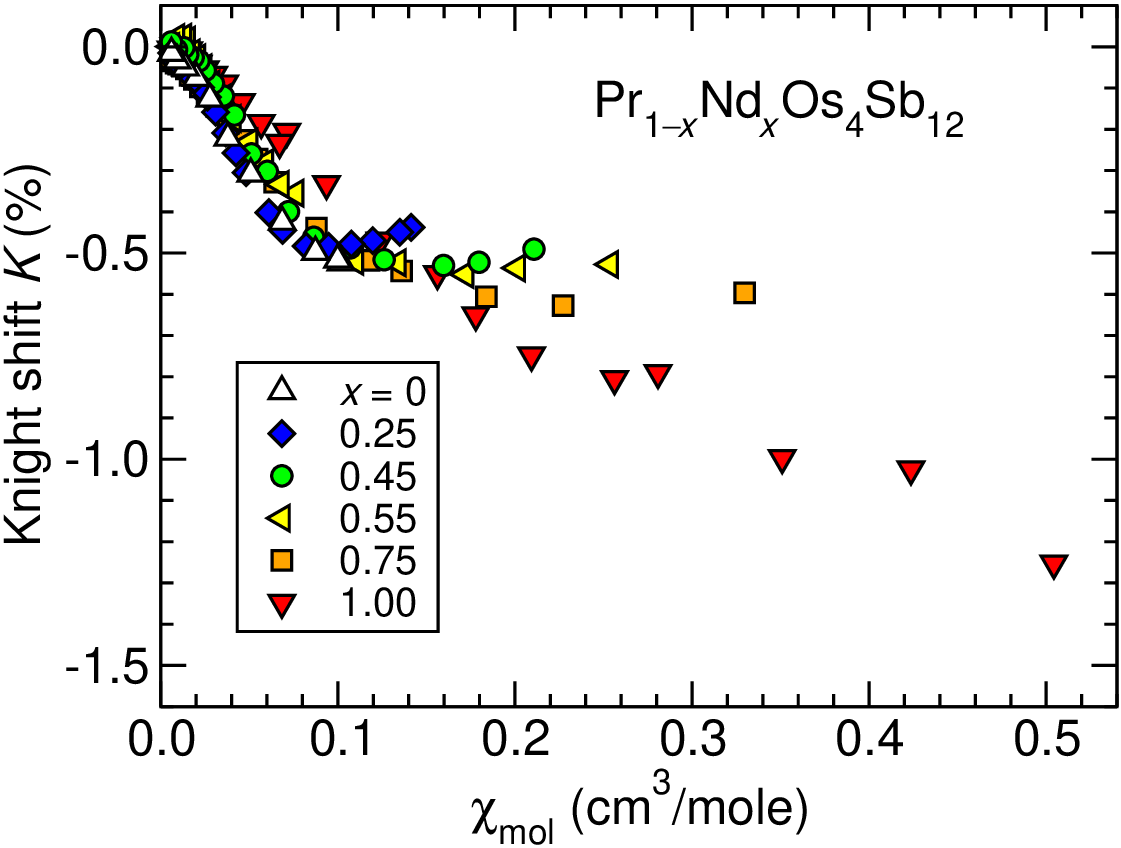}
\caption{\label{fig:PNOSKvschi} (Color online) Clogston-Jaccarino plots of Knight shift~$K$ vs molar susceptibility~$\chi_\mathrm{mol}$ for Pr$_{1-x}$Nd$_x$Os$_4$Sb$_{12}$, where $x = 0$, 0.25, 0.45, 0.55, 0.75, and 1.00. For clarity the error bars are not shown.}
\end{figure}

Clogston-Jaccarino plots for Pr$_{1-x}$Nd$_x$Os$_4$Sb$_{12}$, $x = 0$, where 0.25, 0.45, 0.55, 0.75, and 1.00 are given in Fig.~\ref{fig:PNOSKvschi}. 
For $x \le 0.75$ the behavior is very different from that of NdOs$_4$Sb$_{12}$: at low temperatures $K$ saturates at $-0.5$ to $-0.6$\% with increasing $\chi_\mathrm{mol}$, with the upturn beginning at $\chi_\mathrm{mol} \approx 0.1~\mathrm{cm}^3$/mole. This continues the trend started in PrOs$_4$Sb$_{12}$ [Fig.~\ref{fig:POSK+deltaKvschi}(a)]. 

\begin{figure}[ht] % Fig. 7
\includegraphics[clip=,width=8.6cm]{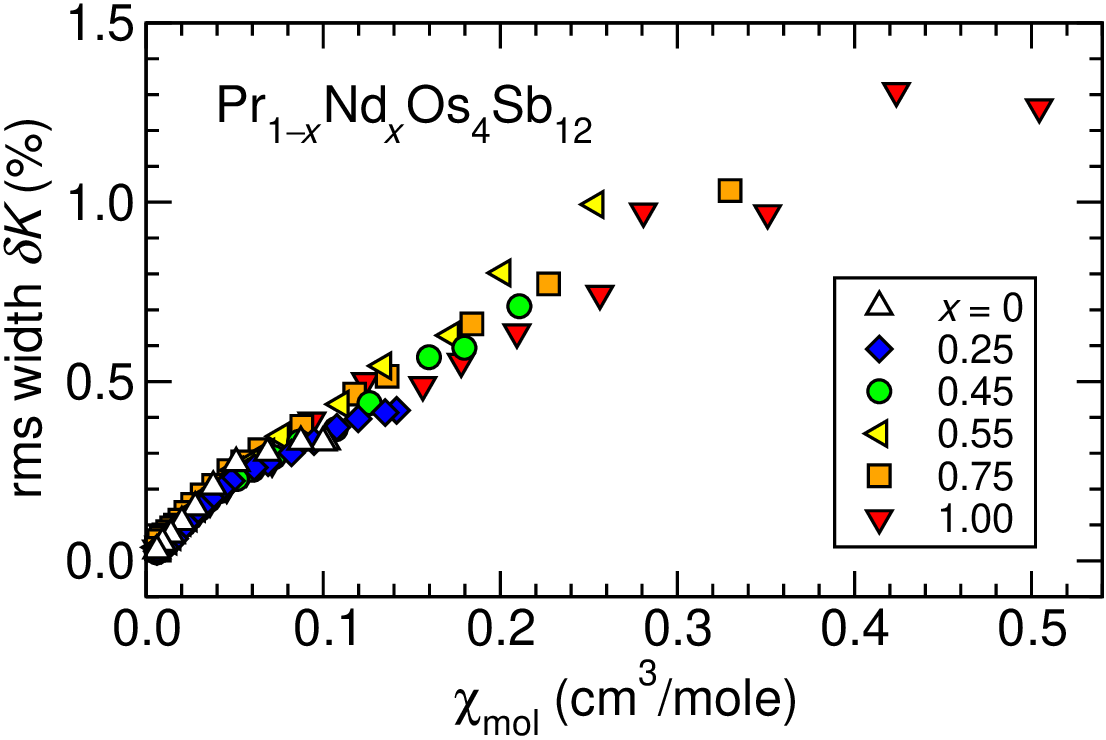}
\caption{\label{fig:PNOSdeltaKvschi} (Color online) rms widths~$\delta K$ of Knight shift distribution vs $\chi_\mathrm{mol}$ for Pr$_{1-x}$Nd$_x$Os$_4$Sb$_{12}$, where $x = 0$, 0.25, 0.45, 0.55, 0.75, and 1.00.}
\end{figure}

The rms widths~$\delta K(\chi_\mathrm{mol})$ for these samples, shown in Fig.~\ref{fig:PNOSdeltaKvschi}, exhibit a slight but discernible change of slope at ${\sim}0.05~\mathrm{cm}^3$/mole, in the region where $K(\chi_\mathrm{mol})$ begins to saturate (Fig.~\ref{fig:PNOSKvschi}). 
Although there is some spread for different $x$ spread at large $\chi_\mathrm{mol}$, the data for all samples, including $x = 1$, are closer to lying on a universal curve than for $K(\chi_\mathrm{mol})$. It is remarkable that for comparable susceptibilities, the widths in the substituted alloys are not much larger than those in the end compounds. Structural disorder is presumably necessary for disorder in Knight shifts, but its origin and magnitude do not seem to play a large role.

\section{\label{sec:disc} DISCUSSION}

We consider the possibility that the nonlinear $K$-$\chi$ relations observed in Pr$_{1-x}$Nd$_x$Os$_4$Sb$_{12}$ for $x < 1$ are due to the differences in hyperfine coupling constants~$A_c^\mathrm{Pr}$ and $A_c^\mathrm{Nd}$ and ionic susceptibilities $\chi^\mathrm{Pr}(x,T)$ and $\chi^\mathrm{Nd}(x,T)$ of Pr$^{3+}$ and Nd$^{3+}$, respectively (per mole ion; we drop the subscript `mol' here and in the following). In this scenario the susceptibility $\chi(x,T)$ and average shift~$K_\mathrm{av}(x,T)$ are given by
\begin{eqnarray} \label{eq:Kavdef} % Eq. (3)
\chi(x,T) & = & (1-x)\chi^\mathrm{Pr}(x,T) + x\chi^\mathrm{Nd}(x,T)\quad \mathrm{and} \nonumber \\
K_\mathrm{av}(x,T) & = & (1-x)A_c^\mathrm{Pr}\chi^\mathrm{Pr}(x,T) + xA_c^\mathrm{Nd}\chi^\mathrm{Nd}(x,T),
\end{eqnarray}
respectively. Ho \textit{et al.}~\cite{HYYD11} argue that Nd doping does not strongly effect the CEF-split electronic structure, so that $\chi^\mathrm{Pr}(x,T)$ is approximately the same in Pr$_{1-x}$Nd$_x$Os$_4$Sb$_{12}$ as in PrOs$_4$Sb$_{12}$. Then the Pr$^{3+}$ contribution to $\chi(x,T)$ of a Nd-doped alloy is just $(1-x)\chi^\mathrm{Pr}(x{=}0,T)$, so that~\cite{HYYD11}
\begin{equation} \label{eq:chiNd} % Eq. (4)
\chi^\mathrm{Nd}(x,T) = \frac{\chi(x,T) - (1-x)\chi^\mathrm{Pr}(0,T)}{x}
\end{equation}
[$\chi^\mathrm{Nd}(x,T) \ne \chi^\mathrm{Nd}(x{=}1,T)$, because Nd-Nd exchange interactions decrease with decreasing $x$]. Then
\begin{eqnarray} \label{eq:Kav} % Eq. (5)
K_\mathrm{av}(x,T) & = & (1-x)\left(A_c^\mathrm{Pr} - A_c^\mathrm{Nd}\right)\chi^\mathrm{Pr}(0,T) \nonumber \\ 
& & +\,A_c^\mathrm{Nd}\chi(x,T)
\end{eqnarray}
from Eqs.~(\ref{eq:Kavdef}) and (\ref{eq:chiNd}). 

\begin{figure}[ht] % Fig. 8
\includegraphics[clip=,width=8.6cm]{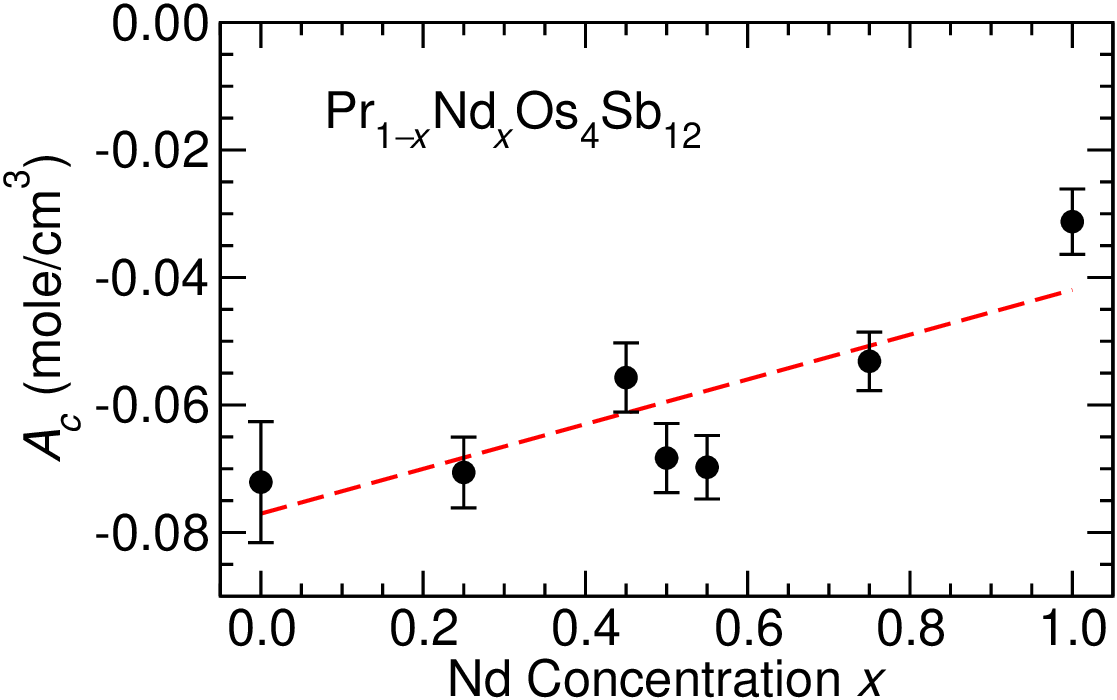}
\caption{\label{fig:PNOSdKdchi} (Color online) Dependence of initial Clogston-Jaccarino plot slopes~$A_c(x) = [dK(\chi,x)/d\chi]_0$ on Nd concentration~$x$ in Pr$_{1-x}$Nd$_x$Os$_4$Sb$_{12}$. Dashed curve: fit of Eq.~(\protect\ref{eq:Acavvsx}) to the data.}
\end{figure}

At high temperatures $\chi^\mathrm{Pr}(x,T)$ and $\chi^\mathrm{Nd}(x,T)$ are Curie-like and yield effective magneton numbers $(p_\mathrm{eff})_i$, $i = \mathrm{Pr}$, Nd, close to the Hund's-rule values~\cite{MFHY06,KTII08}. From Eq.~(\ref{eq:Kavdef}) we then have
\begin{eqnarray} \label{eq:Acavvsx} % Eq. (6)
A_c(x) & = & K_\mathrm{av}(x,T)/\chi(x,T) \nonumber \\
& = & \frac{(1-x)A_c^\mathrm{Pr}(p_\mathrm{eff})_\mathrm{Pr}^2 + xA_c^\mathrm{Nd}(p_\mathrm{eff})_\mathrm{Nd}^2}{(1-x)(p_\mathrm{eff})_\mathrm{Pr}^2 + x(p_\mathrm{eff})_\mathrm{Nd}^2}
\end{eqnarray}
for the initial slope~$A_c(x)$, which is essentially linear in $x$ due to the near equality of $(p_\mathrm{eff})_\mathrm{Pr}$ and $(p_\mathrm{eff})_\mathrm{Nd}$ (ratio 1.011) (Sec.~\ref{sec:results}). Figure~\ref{fig:PNOSdKdchi} shows the Nd concentration dependence of $A_c(x)$ from Clogston-Jaccarino plots for Pr$_{1-x}$Nd$_x$Os$_4$Sb$_{12}$ alloys. 
Equation~(\ref{eq:Acavvsx}) (dashed curve in Fig.~\ref{fig:PNOSdKdchi}) gives a reasonable fit to the data.

\begin{figure}[ht] % Fig. 9
\includegraphics[clip=,width=8.6cm]{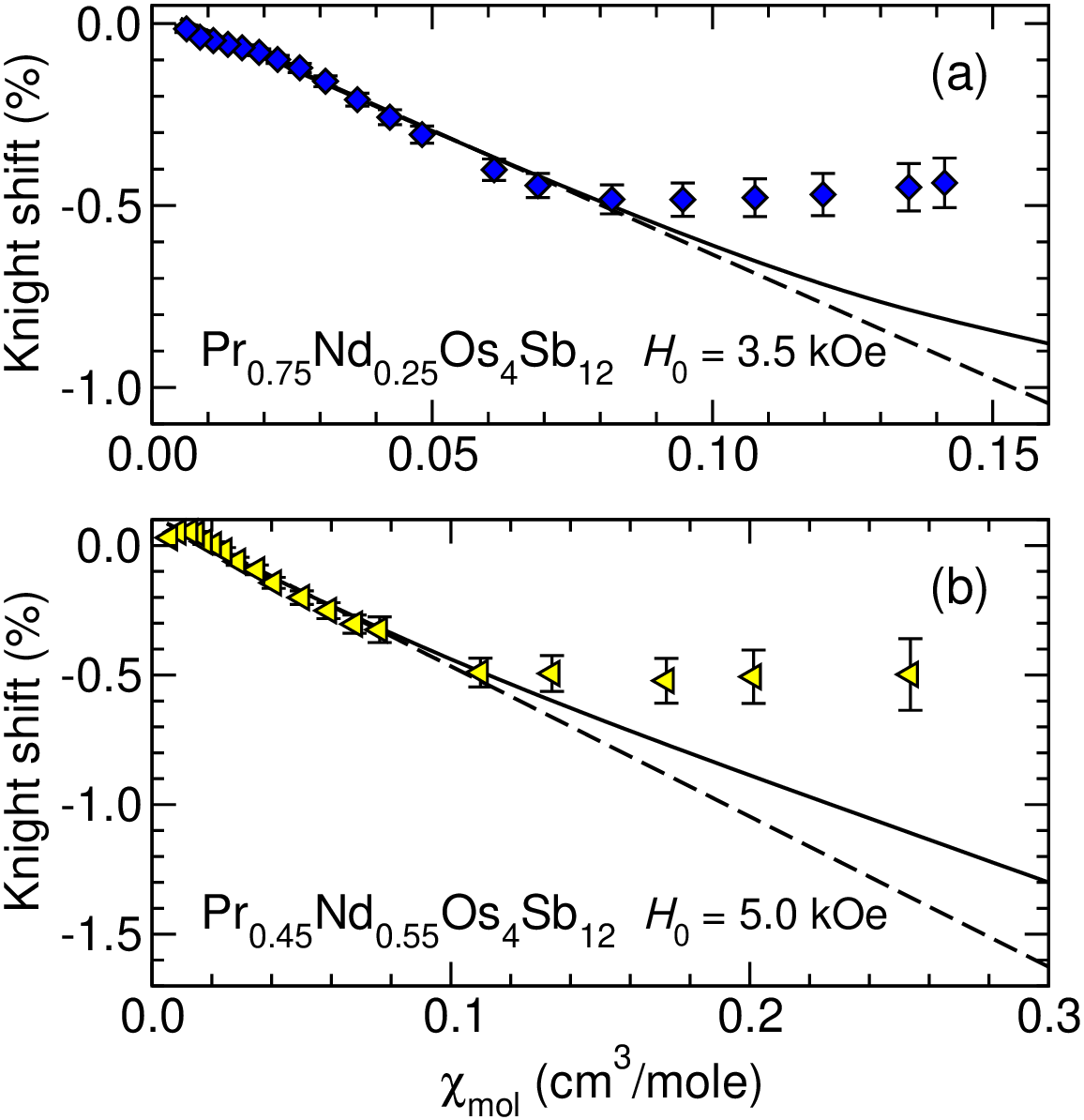}
\caption{\label{fig:PN2575OSKvschi} (Color online) Clogston-Jaccarino plots of Knight shift~$K$ versus molar susceptibility~$\chi_\mathrm{mol}$ for Pr$_{1-x}$Nd$_x$Os$_4$Sb$_{12}$. (a)~$x = 0.25$. (b)~$x = 0.55$. Data (symbols) from Fig.~\protect\ref{fig:PNOSKvschi}. Solid curves: $K_\mathrm{av}$ from Eq.~(\protect\ref{eq:Kav}). Dashed lines: linear fits for $\chi_\mathrm{mol}$ in range~0.01--0.05~cm$^3$/mole.}
\end{figure}

In general Eq.~(\ref{eq:Kav}) yields a nonlinear relation between $K_\mathrm{av}$ and $\chi$. Knight shift data from samples with $x = 0.25$ and 0.55 are compared with $K_\mathrm{av}$ from Eq.~(\ref{eq:Kav}) in the Clogston-Jaccarino plots of Fig.~\ref{fig:PN2575OSKvschi}. 
It can be seen that the slight upward curvature exhibited by $K_\mathrm{av}$ is not nearly enough to account for the observed behavior.

\section{\label{sec:concl} CONCLUSIONS}

It is clear that the simple model of Eq.~(\ref{eq:Kav}) does not capture the behavior of the $\mu^+$ Knight shift in Pr$_{1-x}$Nd$_x$Os$_4$Sb$_{12}$. The saturation of $K(\chi)$, slight for PrOs$_4$Sb$_{12}$ and marked for the Nd-doped alloys (Fig.~\ref{fig:PNOSKvschi}), is evidence for a strong reduction of the hyperfine coupling constant with decreasing temperature. The much smaller nonlinearity of $K(\chi)$ in NdOs$_4$Sb$_{12}$ [Fig.~\ref{fig:NOSK+deltaKvschi}(a)] is probably associated with the ferromagnetic transition at 0.8~K\@. A change of slope but no saturation is observed in $\delta K(\chi)$ (Fig.~\ref{fig:PNOSdeltaKvschi}) for all samples.

It seems difficult to avoid the conclusion that the indirect $\mu^+$-$4f$ ion interaction, which gives the only contribution to the average Knight shift, is strongly affected by the presence of Pr$^{3+}$ ions. The anomalies in both $K(\chi)$ and $\delta K(\chi)$ set in for $\chi(x) \gtrsim 0.05~\mathrm{cm}^3$/mole ($T \lesssim 15$~K) for all samples (Fig.~\ref{fig:PNOSchivsT}). The reduced spontaneous local fields at $\mu^+$ sites reported in I for $0.45 \le x \le 0.75$ are found in magnetic ground states with onset temperatures well below 15~K, and thus might be due to the same coupling-strength reduction.

The Knight-shift saturation sets in at temperatures in the neighborhood of the lowest Pr$^{3+}$ CEF excitation energy in PrOs$_4$Sb$_{12}$~\cite{GOBM04}, suggesting that quadrupolar excitations observed at these energies~\cite{KIKK05} might be involved. Itinerant quadrupolar excitations (quadrupolar excitons) have been suggested as mechanisms for heavy-fermion behavior and Cooper-pair binding in superconducting PrOs$_4$Sb$_{12}$~\cite{MKH03,KMS06,Thal06}. They are not themselves magnetic, but aspherical conduction-electron scattering~\cite{CETF07,ZTF09} together with charge-spin correlations in the conduction band might lead to interference effects in, and suppression of, the $\mu^+$-$4f$ indirect interaction. We note that 10--15~K is also the temperature range below which the susceptibilities begin to diverge with increasing Nd concentration (Fig.~\ref{fig:PNOSchivsT}). This raises the question of whether changes in conduction-electron spin polarization associated with increased Nd-Nd exchange interactions might play a role. But then it would be hard to understand the absence of Knight-shift saturation in NdOs$_4$Sb$_{12}$ [Fig.~\ref{fig:NOSK+deltaKvschi}(a)].

Further investigations are needed of the relation between the anomalous $\mu^+$ Knight shift results reported here and the unique properties of PrOs$_4$Sb$_{12}$ and its alloys, particularly studies of the effect, if any, of quadrupolar excitations on indirect spin-spin interactions.

%\vspace{12pt} 
\begin{acknowledgments}
We are grateful to the Centre for Material and Molecular Sciences, TRIUMF, for facility support during these experiments. Thanks to J.~M. Mackie and B. Samsonuk for assistance with data taking and analysis, and to Y. Aoki and W. Higemoto for useful discussions. This research was supported by the U.S. National Science Foundation, Grants No.~0422674 and No.~0801407 (UC Riverside), No.~0802478 and No.~1206553 (UC San Diego), No.~1104544 (CSU Fresno), and No.~1105380 (CSU Los Angeles), by the U.S. Department of Energy, Grant No.~DE-FG02-04ER46105 (UC San Diego), by the National Natural Science Foundation of China (11204041), the Natural Science Foundation of Shanghai, China (12ZR1401200), and the Research Fund for the Doctoral Program of Higher Education of China (2012007112003) (Shanghai), and by the Japanese MEXT (Hokkaido).
\end{acknowledgments}

%\bibliography{abbrev,CeAl3,CeCu6,comments,frust,HvyFmns,KondoEff,math,muSR,%
%muSRmag,muSRsc,muSRtech,NMR,NMRtech,praseo,PrOsSb,skutt,spinglas,%
%supercon}

%\end{document}

%Merlin.mbs v4.21 2009-07-09.
%

\end{document}